\documentclass[11pt,a4paper]{article}
\usepackage{amsmath}
\usepackage{float}
\usepackage{caption2}

\begin{document}

\title{\textbf{\large{A feasible quantum optical experiment capable of refuting \\
\vspace{-.4cm}noncontextuality for single photons}}}
\vspace{1cm}
\author{Jos\'{e} L.\ Cereceda\thanks{Electronic mail: jl.cereceda@teleline.es} \\
\textit{C/Alto del Le\'{o}n 8, 4A, 28038 Madrid, Spain}}

\date{14 January, 2002}

\maketitle

\begin{abstract}
Elaborating on a previous work by Simon \textit{et al.} [Phys. Rev. Lett. \textbf{85}, 1783 (2000)] we propose a realizable quantum optical single-photon experiment using standard present day technology, capable of discriminating maximally between the predictions of quantum mechanics (QM) and noncontextual hidden variable theories (NCHV). Quantum mechanics predicts a gross violation (up to a factor of 2) of the noncontextual Bell-like inequality associated with the proposed experiment. An actual maximal violation of this inequality would demonstrate (modulo fair sampling) an all-or-nothing type contradiction between QM and NCHV.

\end{abstract}

\vspace{.5cm}

An essential feature inherent in the world picture presented to us by classical physics is that the measured values of any dynamical variable $\mathcal{A}$ ascribed to an individual system are context independent, that is, they do not depend on what other commeasurable variables $\mathcal{B},\mathcal{C},\ldots$ are measured along with it. This central tenet is captured by the so-called noncontextual hidden variable (NCHV) theories which assume that the result of a measurement of $\mathcal{A}$ depends solely on the choice of $\mathcal{A}$ and on the state of the system being measured (the state being fully described by certain (noncontextual) hidden variables). While quantum mechanics (QM) satisfies this noncontextuality condition for the expectation value of the observables, the Bell--Kochen--Specker  theorem [1--3] shows that noncontextuality is incompatible with the predictions that QM makes for \textit{single\/} individual systems, provided the dimension of the associated Hilbert space is greater than two. Experimentally speaking, however, there have been relatively very few experiments testing the validity of the general assumption of noncontextuality, in contrast to the many experiments performed to test Bell's theorem [4] against \textit{local\/} hidden variables (LHV) [5]. Indeed, to our knowledge, the recent two experiments performed by Michler \textit{et al.} [6] are the first explicitly implementing a statistical test of NCHV versus QM. The second one of these experiments [6] is an `event ready' test of a Bell-like inequality involving only one particle, derived from the noncontextuality assumption (see below for a brief discussion on this experiment). Also recently, in a very interesting paper Simon \textit{et al.} [7], inspired by an argument by Cabello and Garc\'{\i}a-Alcaine [8], developed a simple experimental scheme allowing a nonstatistical test of noncontextuality. This scheme works with single particles, and uses both their spin and translational degrees of freedom.

In this paper we present a practical quantum optical method for Simon \textit{et al.}'s scheme which can be implemented with a suitable source of polarized single photons and a set of polarizing beam splitters and photodetectors. The original scheme in [7] exhibits a direct contradiction between NCHV and QM for \textit{every\/} run of the experiment, so that, ideally, it would suffice a single particle to either prove or disprove noncontextuality. In practice, however, due to the imperfect detection efficiency and other nonidealities, it is not possible to attain the perfect correlations on which this all-or-nothing type contradiction relies. In fact, all we can do in a real experiment is to execute a sufficiently large number of runs in order to find out the experimentally significant ensemble averages. As we will see, an actual realization of the scheme in [7] requires determining the ensemble average of three appropriate observables for single particles. It turns out that the assumption of noncontextuality implies an upper bound on the absolute value of a certain linear combination of these three ensemble averages (see Eq.\ (7) below), whereas the quantum-mechanical predictions can exceed this bound by a factor as large as 2 (which is the maximum possible theoretical violation of the inequality at issue). An empirical violation of this one-particle Bell-like inequality would then invalidate (modulo fair sampling) the noncontextuality assumption. Furthermore, a violation of inequality (7) by a factor approaching 2 would mean that we are refuting noncontextuality for \textit{every\/} particle of the ensemble (again, modulo fair sampling).  

Before describing our experimental proposal for Simon \textit{et al.}'s scheme we briefly recall the logical contradiction between NCHV theories and QM entailed by the argument in [7] (see also [8]). Consider four two-valued observables $Z_1$, $X_1$, $Z_2$, and $X_2$ ascribed to a single individual system, the measurement results for each observable being denoted by $\pm1$. It is assumed that, in a NCHV theory, the above observables have predefined noncontextual values of either $+1$ or $-1$, which are denoted as $v(Z_1)$, $v(X_1)$, $v(Z_2)$, and $v(X_2)$, respectively. Consider now an ensemble of systems for which the results of $Z_1$ and $Z_2$ are always found to be equal to each other, and the same for the results of $X_1$ and $X_2$. Then it can be shown [7] that, for such an ensemble, NCHV predicts that $v(Z_1 X_2) = v(X_1 Z_2)$, where, by definition, for a NCHV theory $v(Z_1 X_2)$ is equal to the product $v(Z_1)v(X_2)$. (Note that the considered ensemble can then equivalently be defined by the condition $v(Z_1 Z_2)=v(X_1 X_2)=1$.) On the other hand, for the corresponding  quantum mechanical observables $Z_1$, $X_1$, $Z_2$, and $X_2$, one can find a two-qubit state $|\Psi_1\rangle$ which is a joint eigenstate of the (commuting) product observables $Z_1 Z_2$ and $X_1 X_2$, with eigenvalue $+1$ for both of them. Moreover, for this two-qubit state QM predicts that the measured value of the observable $Z_1 X_2$ will always be opposite to the measured value of the (commuting) observable $X_1 Z_2$, in direct contradiction with the above NCHV prediction according to which $v(Z_1 X_2) = v(X_1 Z_2)$. Mathematically, this contradiction between QM and NCHV stems from the fact that the following set of equations entailed by QM,
\begin{align}
Z_1 Z_2 |\Psi_1\rangle = & \,\, |\Psi_1\rangle,  \nonumber  \\
X_1 X_2 |\Psi_1\rangle = & \,\, |\Psi_1\rangle,   \\
Z_1 X_2 |\Psi_1\rangle = & \, - X_1 Z_2|\Psi_1\rangle,  \nonumber
\end{align}
cannot be fulfilled by a NCHV theory. Indeed, in order for a NCHV description to be consistent with Eqs.\ (1), it is necessary that
\begin{align}
v(Z_1) \, v(Z_2) = & \, +1,    \tag{2a}   \\
v(X_1) \, v(X_2) = & \, +1,    \tag{2b}   \\
v(Z_1) \, v(X_2) = & \, -v(X_1) \, v(Z_2),  \tag{2c}
\setcounter{equation}{2}
\end{align}
where $v(Z_1)$ in Eq.\ (2a) assumes the same value as $v(Z_1)$ in Eq.\ (2c), etc. However, it is impossible to assign values, either $+1$ or $-1$, that satisfy the constraints in Eqs.\ (2a)-(2c), as can easily be verified by taking the product of such equations.

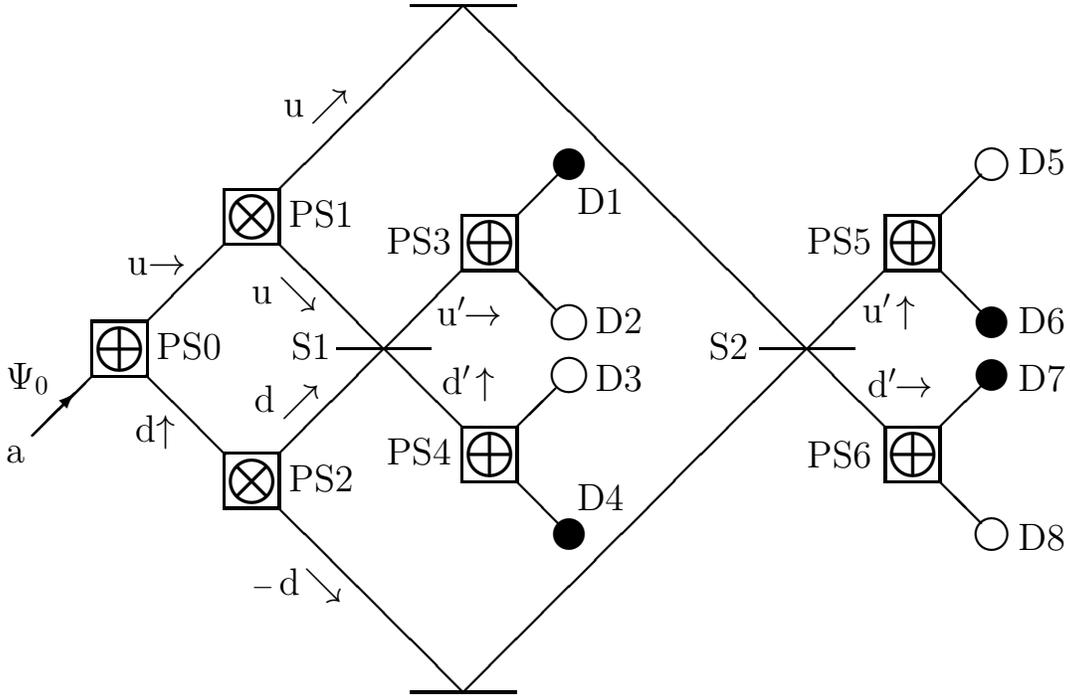
\begin{figure}[ttt]
    
\begin{picture}(400,225)(25,60)
\thicklines

\put(47,135){\Large{$\Psi_0$}}
\put(47,107){\Large{a}}
\put(57,117){\line(1,1){23}}
\put(57,117){\vector(1,1){16}}

\put(80,140){\framebox(20,20){\LARGE{$\bigoplus$}}}\put(104,146){\Large{PS0}}
\put(100,160){\line(1,1){30}}\put(93,178){\Large{u}$\rightarrow$}
\put(130,190){\framebox(20,20){\LARGE{$\bigotimes$}}}\put(154,196){\Large{PS1}}
              \put(152,238){\Large{u}\,$\nearrow$}
              \put(140,167){\Large{u}\,$\searrow$}
\put(150,210){\line(1,1){70}}
\put(220,280){\line(1,-1){160}}

\put(100,140){\line(1,-1){30}}\put(96,115){\Large{d}$\uparrow$}
\put(130,90){\framebox(20,20){\LARGE{$\bigotimes$}}}\put(154,96){\Large{PS2}}
              \put(140,57){$-\,$\Large{$\text{d}\!\searrow$}}
              \put(141,126){\Large{d}\,$\nearrow$}
\put(150,90){\line(1,-1){70}}
\put(220,20){\line(1,1){160}}

\put(150,190){\line(1,-1){70}}
\put(150,110){\line(1,1){70}}
\put(220,180){\framebox(20,20){\LARGE{$\bigoplus$}}}\put(191,186){\Large{PS3}}
\put(220,100){\framebox(20,20){\LARGE{$\bigoplus$}}}\put(191,106){\Large{PS4}}

\put(240,200){\line(1,1){20}}
\put(240,180){\line(1,-1){15}}
\put(240,120){\line(1,1){15}}
\put(240,100){\line(1,-1){20}}
\put(260,220){\circle*{12}}\put(263,201){\Large{D1}}
\put(260,160){\circle{13}}\put(270,156){\Large{D2}}
\put(260,140){\circle{13}}\put(270,134){\Large{D3}}
\put(260,80){\circle*{12}}\put(263,89){\Large{D4}}

\put(380,180){\framebox(20,20){\LARGE{$\bigoplus$}}}\put(350,186){\Large{PS5}}
\put(380,100){\framebox(20,20){\LARGE{$\bigoplus$}}}\put(350,105){\Large{PS6}}
\put(400,200){\line(1,1){16}}
\put(400,180){\line(1,-1){20}}
\put(400,120){\line(1,1){20}}
\put(400,100){\line(1,-1){16}}
\put(420,220){\circle{12}}\put(430,216){\Large{D5}}
\put(420,160){\circle*{12}}\put(430,156){\Large{D6}}
\put(420,140){\circle*{12}}\put(430,134){\Large{D7}}
\put(420,80){\circle{12}}\put(430,74){\Large{D8}}

\put(200,280){\line(1,0){40}}
\put(200,20){\line(1,0){40}}
\put(172,150){\line(1,0){36}} \put(155,146){\Large{S1}}
              \put(210,159){\Large{$\text{u}^{\prime}\!\!\rightarrow$}}
              \put(212,133){\Large{$\text{d}^{\prime}\!\uparrow$}}
\put(332,150){\line(1,0){36}} \put(313,146){\Large{S2}}
              \put(371,160){\Large{$\text{u}^{\prime}\!\uparrow$}}
              \put(373,132){\Large{$\text{d}^{\prime}\!\!\rightarrow$}}

\end{picture}
\setcaptionmargin{1cm}
\captionstyle{centerlast}
\renewcommand{\figurename}{Fig.}
\renewcommand{\captionlabeldelim}{.~}
\vspace{1.7cm}
\caption{\small{Proposed experimental setup effecting the original scheme in [7]. Quantum mechanics predicts that the single photons (prepared in a suitable initial state $|\Psi_0\rangle$) can be detected only at the black-coloured detectors, while NCHV predicts detections at the complementary set of white-coloured ones (see the main text for details).}}
\vspace{.2cm}
\end{figure}

Let us now analyse the proposed interferometric arrangement depicted in Fig.~1 realizing the experimental scheme in [7]. A single photon linearly polarized along $+45^{\circ}$ enters the interferometer from the left in the spatial mode $a$. The polarizing beam splitter PS0 transmits the horizontal ($0^{\circ}$) polarization component of the incoming photon and reflects the vertical ($90^{\circ}$) one, so that a photon polarized along the horizontal (vertical) direction will with certainty exit via the upper (lower) output port of PS0, labelled as $\text{u}\!\rightarrow$ ($\text{d}\!\uparrow$) in Fig.~1. PS1 and PS2 are polarizing beam splitters with PS1 (PS2) transmitting the $+45^{\circ}$ ($-45^{\circ}$) polarization component while it reflects the $-45^{\circ}$ ($+45^{\circ}$) component, the diagonal polarization basis $\{ |+\nolinebreak 45^{\circ}\rangle, |-45^{\circ}\rangle \} \equiv \{ |\nearrow\rangle, |\searrow\rangle \}$ being related to the rectilinear polarization basis $\{ |0^{\circ}\rangle, |90^{\circ}\rangle \} \equiv \{ |\rightarrow\rangle, |\uparrow\rangle \}$ by
\begin{align}
|\nearrow\rangle &= (1/\sqrt{2}) (|\rightarrow\rangle + |\uparrow\rangle ),  \nonumber  \\[-.33cm]
&  \\[-.33cm]
|\searrow\rangle &= (1/\sqrt{2}) (|\rightarrow\rangle - |\uparrow\rangle ).  \nonumber
\end{align}
S1 and S2 are nonpolarizing 50--50 beam splitters with input modes $u$ and $d$ and output modes $u^{\prime}$ and $d^{\prime}$, such that the translational state of a photon entering either S1 or S2 via one of its input modes will undergo the following unitary transformation
\begin{align} 
|u\rangle \to & \, (1/\sqrt{2}) (|u^{\prime}\rangle + |d^{\prime}\rangle ),  \nonumber  \\[-.33cm]
&   \\[-.33cm]
|d\rangle \to & \, (1/\sqrt{2}) (|u^{\prime}\rangle - |d^{\prime}\rangle ),  \nonumber
\end{align}
which is independent of the polarization state of the photon. Finally, PS3, PS4, PS5, and PS6 are polarizing beam splitters which measure the polarization of the incoming photon in the rectilinear basis $\{ |\rightarrow\rangle, |\uparrow\rangle \}$, with the photon ending up in one of the photodetectors $\text{D1},\text{D2},\ldots,\text{D8}$.

The observables to be considered in our interferometric setup are
\begin{align}
Z_1 = & \,\, |u\rangle\langle u| - |d\rangle\langle d|, \nonumber  \\
X_1 = & \,\, |u^{\prime}\rangle\langle u^{\prime}| - 
|d^{\prime}\rangle\langle d^{\prime}|, \nonumber  \\[-.33cm]
&  \\[-.33cm]
Z_2 = & \,\, |\rightarrow\rangle\langle\rightarrow| - |\uparrow\rangle\langle\uparrow|, \nonumber \\
X_2 = & \,\, |\nearrow\rangle\langle\nearrow| - |\searrow\rangle\langle\searrow|,  \nonumber
\end{align}
where $Z_1$ and $X_1$ ($Z_2$ and $X_2$) involve the translational (polarization) degree of freedom of a single photon propagating inside the interferometer. It is easily verified that every single photon prepared in the initial state $|\Psi_0\rangle = |a\rangle |\nearrow\rangle$ will always yield $+1$ for the measured values of the product observables $Z_1 Z_2$ and $X_1 X_2$. This follows from the fact that, upon interacting with PS0,  the state $|\Psi_0\rangle$ transforms into $|\Psi_1 \rangle = \tfrac{1}{\sqrt{2}}(|u\rangle|\rightarrow\rangle +|d\rangle|\uparrow\rangle) = \linebreak \tfrac{1}{\sqrt{2}}(|u^{\prime}\rangle|\nearrow\rangle +
|d^{\prime}\rangle|\searrow\rangle)$. Moreover, the detection of a photon at any given detector corresponds to a \textit{simultaneous\/} measurement of the product observables $Z_1 X_2$ and $X_1 Z_2$. To see this, consider the case when a photon is detected by D1, say. This immediately means that, for the detected photon, $X_1 =+1$ and $Z_2 =+1$, and then $X_1 Z_2 =+1$ (see Fig.~1). On the other hand, the state of a photon just before reaching S1 consists of a superposition of the states $|u\rangle|\searrow\rangle$ and $|d\rangle|\nearrow\rangle$ so that a measurement of $Z_1 X_2$ on this photon must necessarily give the result $Z_1 X_2=-1$. In Table~1 we display the values of $Z_1 X_2$ and $X_1 Z_2$, as revealed by the detection of the photon at the various detectors $\text{D1},\text{D2},\ldots,\text{D8}$. We note that, interestingly, the detection of a photon at any given detector provides the value of the product observable $Z_1 X_2$ as a whole, without revealing any information about the values of $Z_1$ and $X_2$ separately.
Furthermore we should mention that, as pointed out in [7], the observables $Z_1$, $X_1$, $Z_2$, and $X_2$ appear in two different contexts, one involving the measurement of both $Z_1Z_2$ and $X_1X_2$, and the other involving the measurement of both $Z_1X_2$ and $X_1Z_2$. The setup of Fig.~1 performs a joint measurement of $Z_1X_2$ and $X_1Z_2$. The observables $Z_1Z_2$ and $X_1X_2$ could be measured by means of an auxiliary device not shown in Fig.~1.

\begin{table}[t]
\begin{center}
\begin{tabular}{c|cccccccc}
& D1 & D2& D3 & D4 & D5 & D6 & D7 & D8 \\  \hline
$Z_1 X_2$ & $-1$ & $-1$ & $-1$ & $-1$ & $+1$ & $+1$ & $+1$ & $+1$ \\
$X_1 Z_2$ & $+1$ & $-1$ & $-1$ & $+1$ & $+1$ & $-1$ & $-1$ & $+1$ \\
$Z_1 X_2 \cdot X_1 Z_2$ & $-1$ & $+1$ & $+1$ & $-1$ & $+1$ & $-1$ & $-1$ & $+1$
\end{tabular}
\end{center}
\setcaptionmargin{1cm}
\captionstyle{centerlast}
\vspace{-2mm}
\caption{\small{Results of the joint measurement of $Z_1X_2$ and $X_1Z_2$, which is completed when the photon is detected at one of detectors $\text{D1},\text{D2},\ldots,\text{D8}$. Also shown is the resulting value of the observable $Z_1 X_2 \cdot X_1 Z_2$ at each of the detectors.}}
\vspace{.2cm}
\end{table}

Using the relations (3) and (4), it is straightforward to see that, according to QM, a photon entering the interferometer in the state $|\Psi_0\rangle$ can reach only detectors D1, D4, D6, and D7 (each with probability $\tfrac{1}{4}$). Note that such detectors entail opposite values of $Z_1 X_2$ and $X_1 Z_2$. On the other hand, according to a NCHV theory, a photon fulfilling the property $Z_1 Z_2 = X_1 X_2 =+1$ must end up at one of the detectors D2, D3, D5, or D8. So, for example, a photon existing in a NCHV state defined by $v(Z_1)=v(Z_2)=-1$ and $v(X_1)=v(X_2)=-1$ would end up at detector D8. So, theoretically, the interferometer setup of Fig.~1 exhibits conflicting predictions between QM and NCHV at a nonstatistical level, that is, for every single photon prepared in the initial state $|\Psi_0\rangle$.

Actually, however, any real experiment deals with a number of unavoidable imperfections and undesirable effects. In particular, for the kind of experimental setup we are considering, the obtainable detection/collection efficiency (typically less than 10\%; see, for example, [6]) means that only a relatively small fraction of the photons emitted by the source are detected. Moreover, it may happen that a photon is registered by a detector other than D1, D4, D6, or D7, due, for example, to an imperfect initial state preparation. So we need to average the measurement results obtained in a large number of runs in order to get experimentally meaningful quantities (the validity of the fair sampling assumption is taken for granted).

Bearing this in mind, we now consider the following combination of product observables
\begin{equation}
C = I + Z_1 Z_2 + X_1 X_2 - Z_1 X_2 \cdot X_1 Z_2 \, ,
\end{equation}
where $I$ is the identity operator, and $Z_1X_2 \cdot X_1Z_2$ denotes the product of $Z_1X_2$ and $X_1Z_2$. In a NCHV theory $C$ has the value $C_{\text{NCHV}}=1+v(Z_1)v(Z_2)+v(X_1)v(X_2)-v(Z_1)v(X_2)v(X_1)v(Z_2)$, where, for example, $v(Z_1)$ in both the second and fourth terms assume the same value. Now, as each $v(Z_1)$, $v(Z_2)$, etc.\ is equal to either $+1$ or $-1$, one can readily check that $C_{\text{NCHV}}=\pm2$. Thus, taking the ensemble average of $Z_1Z_2$, $X_1X_2$, and $Z_1X_2 \cdot X_1Z_2$, we arrive at the following inequality which must necessarily be fulfilled by any NCHV theory
\begin{equation}
|1+\langle Z_1Z_2 \rangle + \langle X_1X_2 \rangle - \langle Z_1X_2 \cdot X_1Z_2 \rangle | \leq 2 \, .
\end{equation}
This Bell-like inequality for single particles can be tested using the experimental arrangement of Fig.~1. Indeed, registering the counts at detectors $\text{D1},\text{D2},\ldots,\text{D8}$, and averaging over a sufficiently large number of counts gives (via the value assignment of Table~1) the quantity $\langle Z_1X_2 \cdot X_1Z_2\rangle$. Thus, provided with the values of $\langle Z_1Z_2\rangle$ and $\langle X_1X_2\rangle$ which are obtained in a separate experiment, one can experimentally determine the left-hand side of inequality (7). For the subensemble of photons detected at any one of detectors $\text{D1},\text{D2},\ldots,\text{D8}$, NCHV imposes an upper bound on $\langle C \rangle$ equal to ${\langle C_{\text{NCHV}}\rangle}^{\text{max}} = [1+\langle Z_1Z_2 \rangle + \langle X_1X_2 \rangle - {\langle Z_1X_2 \cdot X_1Z_2 \rangle]}_{\text{NCHV}}^{\text{max}} = 1+1+1-1=2$, in accordance with (7). Quantum mechanics, on the other hand, predicts an upper bound of ${\langle C_{\text{QM}}\rangle}^{\text{max}} = [1+\langle Z_1Z_2 \rangle + \langle X_1X_2 \rangle - {\langle Z_1X_2 \cdot X_1Z_2 \rangle]}_{\text{QM}}^{\text{max}} = 1+1+1+1=4$, which violates maximally the inequality (7) by a factor of 2.

An empirical violation of the inequality (7) would disprove the class of NCHV theories which assume predefined noncontextual values of the observables $Z_1$, $X_1$, $Z_2$, and $X_2$. For the setup of Fig.~1, the maximal violation of the inequality (7) is attained when each member of the subensemble of detected photons is registered at either D1, D4, D6, or D7, which results in a direct contradiction between QM and NCHV for every photon of the original ensemble described by the state $|\Psi_0\rangle$, provided fair sampling holds good. It is worth noting, incidentally, that this all-or-nothing type contradiction between QM and NCHV has been obtained for a quantum system pertaining to a four-dimensional (tensor product) Hilbert space. This is to be compared with the case of Bell's theorem, where it is necessary to consider at least \textit{three\/} spin-$\tfrac{1}{2}$ particles (described in a (tensor product) Hilbert space of dimension eight) in order to get an all-or-nothing type contradiction between QM and LHV [9,10].

The predicted quantum-mechanical violation of the inequality (7) associated with the experimental scheme of Fig.~1, is greater than that obtained in previous schemes involving single particles described in terms of a four-dimensional Hilbert space [6,11] (for completeness see also [12], where noncontextual Bell-type inequalities are derived for single particles of high spin). For example, the inequality tested in the relevant second experiment reported in [6] is violated by QM by a factor up to $\sqrt{2}$. This latter inequality involves correlation functions for pairs of observables ascribed to a single particle (photon 1). The determination of such correlation functions for photon 1, however, requires (for each run of the experiment) the previous detection of a second particle (photon 2) which is in turn correlated with photon 1, the registration of photon 2 at a given detector serving as a trigger event for the performance of the experiment on photon 1. Thus the overall realization of the second experiment in [6] does involve an ensemble of entangled photon pairs in a necessary way, and hence it lacks the simplicity of our proposed experiment which makes use of an ensemble of polarized single photons.

In conclusion, in this paper we have presented a feasible method for the original scheme of Simon \textit{et al.} [7] which can be implemented with present day optical technology. Specifically, the interferometer setup sketched in Fig.~1 offers the basis for performing a real single-photon experiment capable of discriminating maximally (via the testable Bell-like inequality (7)) between the predictions of QM and NCHV.

\newpage

\center


\begin{thebibliography}{99}

\bibitem{} J. S. Bell, Rev. Mod. Phys. \textbf{38}, 447 (1966)

\bibitem{} S. Kochen and E. P. Specker, J. Math. Mech. \textbf{17}, 59 (1967).

\bibitem{} N. D. Mermin, Phys. Rev. Lett. \textbf{65}, 3373 (1990); N. D. Mermin, Rev. Mod. Phys. \textbf{65}, 803 (1993).

\bibitem{} J. S. Bell, Physics \textbf{1}, 195 (1964).

\bibitem{} A nonexhaustive list of relevant experiments testing Bell's inequality is: S. J. Freedman and J. F. Clauser, Phys. Rev. Lett. \textbf{28}, 938 (1972); A. Aspect, J. Dalibard, and G. Roger, Phys. Rev. Lett. \textbf{49}, 1804 (1982); Y. H. Shih and C. O. Alley, Phys. Rev. Lett. \textbf{61}, 2921 (1988); J. G. Rarity and P. R. Tapster, Phys. Rev. Lett. \textbf{64}, 2495 (1990); P. G. Kwiat, K. Mattle, H. Weinfurter, A. Zeilinger, A. V. Sergienko, and Y. H. Shih, Phys. Rev. Lett. \textbf{75}, 4337 (1995); G. Weihs, T. Jennewein, C. Simon, H. Weinfurter, and A. Zeilinger, Phys. Rev. Lett. \textbf{81}, 5039 (1998); M. A. Rowe, D. Kielpinski, V. Meyer, C. A. Sackett, W. M. Itano, C. Monroe, and D. J. Wineland, Nature \textbf{409}, 791 (2001).

\bibitem{} M. Michler, H. Weinfurter, and M. \.{Z}ukowski, Phys. Rev. Lett. \textbf{84}, 5457 (2000).

\bibitem{} C. Simon, M. \.{Z}ukowski, H. Weinfurter, and A. Zeilinger, Phys. Rev. Lett. \textbf{85}, 1783 (2000).

\bibitem{} A. Cabello and G. Garc\'{\i}a-Alcaine, Phys. Rev. Lett. \textbf{80}, 1797 (1998).

\bibitem{} D. M. Greenberger, M. A. Horne, and A. Zeilinger, in \textit{Bell's Theorem, Quantum Theory, and Conceptions of the Universe,} edited by M. Kafatos (Kluwer, Dordrecht, 1989), pp.\ 69-72; D. M. Greenberger, M. A. Horne, A. Shimony, and A. Zeilinger, Am. J. Phys. \textbf{58}, 1131 (1990); N. D. Mermin, Am. J. Phys. \textbf{58}, 731 (1990).

\bibitem{} G. Nistic\`{o}, Phys. Lett. A \textbf{281}, 273 (2001).

\bibitem{} S. Basu, S. Bandyopadhyay, G. Kar, and D. Home, Phys. Lett. A \textbf{279}, 281 (2001).

\bibitem{} S. M. Roy and V. Singh, Phys. Rev. A \textbf{48}, 3379 (1993).


\end{thebibliography}
\end{document}